%% file: metrid18.tex
\documentclass[submission, copyright,creativecommons]{eptcs}
\providecommand{\event}{MeTRiD 2018} 
\usepackage{underscore}           
\usepackage{color}
\usepackage{xspace}
\usepackage[T1]{fontenc}
\usepackage{todonotes}
\usepackage{comment}
\usepackage{appendix}
\usepackage{versions}
\usepackage{url}
\usepackage{hyperref}
\usepackage{cleveref}
\usepackage{amsmath}
\usepackage{array}
\usepackage{jmllistings,multicol}
\usepackage{latexsym,amssymb}

\input defs-sac18.tex

\newcommand{\fracfun}{p}
\newcommand{\fracminus}{\mathop{\dot{\smash{-}}}}

\newcommand{\blind}[2]{
\ifdefined\blindon
    #1\xspace
\else
   #2\xspace
\fi
}

\title{Verification of Shared-Reading Synchronisers}
\author{
\blind{blind review}
{
Afshin Amighi
\institute{Hogeschool Rotterdam\thanks{The research is done while working at University of Twente.}\\ Rotterdam, The Netherlands}
\email{a.amighi@hr.nl}
\and
Marieke Huisman
\institute{University of Twente\\ Enschede, The Netherlands}
\email{m.huisman@utwente.nl}
\and
Stefan Blom
\institute{BetterBe$^{*}$ \\ Enschede, The Netherlands}
\email{sblom@betterbe.com}
}
}
\def\titlerunning{Verification of Shared-Reading Synchronisers}
\def\authorrunning{\blind{blind review}{A. Amighi, M. Huisman \& S. Blom}}
\begin{document}
\maketitle

\begin{abstract}
Synchronisation classes are an important building block for shared
memory concurrent programs. Thus to reason about such programs, it is important to be able to verify the
implementation of these synchronisation classes, considering
atomic operations as the synchronisation
primitives on which the implementations are built. 
For synchronisation classes controlling exclusive access to a shared
resource, such as locks, a technique has been proposed to reason about their behaviour. 
This paper proposes a technique to verify implementations of both \emph{exclusive access} and
\emph{shared-reading} synchronisers. We use permission-based Separation Logic to describe the behaviour of the main
atomic operations, and the basis for our technique is formed by a specification for
class \incode{AtomicInteger},  which is commonly used to implement
synchronisation classes in  \incode{java.util.concurrent}. To
demonstrate the applicability of our approach, we mechanically verify
the implementation of various synchronisation classes like
\incode{Semaphore}, \incode{CountDownLatch} and
\incode{Lock}. 

\end{abstract}


\section{Introduction}

As our society is increasingly becoming more digital, there
is an urgent need for techniques that can improve the performance of
software. Concurrency is commonly used to achieve this goal, as it
allows to split bigger tasks into multiple smaller tasks, which can be
executed simultaneously. 
This has many advantages, but also a major disadvantage, namely that developing concurrent software is error-prone, as one has to keep track of how all threads can interact with each other. 
Therefore, we need techniques that allow to specify and verify the behaviour of concurrent programs.

In this paper, we consider this problem. We focus in particular on
shared memory concurrent programs, where multiple threads interact and
communicate via a common, shared memory. One of the main building
blocks of such programs are synchronisation classes that control the
access to a shared memory. We distinguish between two important classes of
synchronisers: \emph{exclusive accesses} synchronisers, such as locks, that
ensure that always at most one thread at a time can access a shared
memory location, and \emph{shared-reading} synchronisers, such as semaphores,
that also allow multiple threads to read the same shared location
simultaneously. The concurrency API
\incode{java.util.concurrent} (JUC) provides several variations of both
kinds of synchronisers, typically implemented on top of the \incode{AtomicInteger} class.

We proposed a technique to specify and verify
\emph{exclusive access} synchronisers, such as \incode{Lock}, using
permission-based Separation Logic~\cite{AmighiBH14}. 
Our paper extends this approach to cover also
\emph{shared-reading} synchronisers. 
The original approach identifies two main components that make up the
specification of a
synchroniser: (1) the value of the atomic variable, i.e. the atomic
state, and (2) the views of the participating threads on the atomic
state, i.e.  the latest value that a thread remembers from the atomic state. 
The client program using the synchroniser specifies a synchronisation
protocol that captures the roles of the threads and a resource
invariant describing the shared memory location protected by the
synchroniser. 

In this paper, we make this approach more fine-grained, allowing a
thread to obtain only a read permission to the shared memory
location. We derive new specifications for the atomic operations that
capture the possibility of obtaining both exclusive and partial access, and
combine these into a new contract for the class
\incode{AtomicInteger}. 

Applicability of the approach is demonstrated by discussing the verification of several commonly used synchronisers: \incode{Semaphore}, \incode{CountDownLatch} and \incode{Lock}.
All examples are mechanically verified using our VerCors tool-set~\cite{BlomH14}. 

The paper is structured as follows:~\cref{sec:background} briefly
introduces permission-based Separation Logic. \cref{sec:synchronisers}
discusses several implementations of typical shared-reading
synchronisers. \cref{sec:specifications} derives the specifications
for the three main atomic operations, using permission-based
Separation Logic. Then, \cref{sec:contracts} combines this into a
contract for \incode{AtomicInteger}, and \cref{sec:verification} shows
how this is used to verify the synchronisation
classes. Finally,~\cref{sec:conclusion} concludes the paper, and discusses related work.

\section{Background}\label{sec:background}

This section briefly explains permission-based Separation Logic (PBSL)~\cite{BornatCOP05} and its role in reasoning about concurrent programs. 
Concurrent Separation Logic (CSL)~\cite{OHearn07}, which is an extension of SL~\cite{Reynolds02}, is a Hoare-style program logic to reason about \emph{multi-threaded} programs. 
In addition to the predicates and operators from first order logic, CSL uses two new constructs in the specifications: 
(1) The \emph{points-to} predicate: $\pointsto e v$ describes that the location of the heap addressed by $e$ is pointing to a location that contains the value $v$, and
(2) The \emph{$\SLAnd$-conjunction} operator: $\phi \SLAnd \psi$ expresses that predicates $\phi$ and $\psi$ hold for two disjoint parts of the heap.
We use $[e]$ to denote the contents of the heap at location $e$, and we use $\pointsto e -$ to indicate that the precise contents stored at location $[e]$ is not important. 
Predicates $P$ and $Q$ of a Hoare-triple $\hoaretriple{P}{C}{Q}$ in CSL are predicates on the state where the state is a pair of the store and the heap. 
The key point of verification using CSL is the \emph{ownership} concept. 
In the verification of $C$ if its precondition $P$ asserts $\pointsto e v$, it is assumed that the executing thread $t$ has the full ownership of $e$. This means that no other thread can interfere with $t$ to update $e$, unless $t$ transfers the ownership of $e$ to another program. 

O'Hearn developed required rules to reason about threads exchanging \emph{exclusive} ownership of a memory location through a synchronisation construct~\cite{OHearn07}. In the rules related to shared memory, the shared state is specified by a \emph{resource invariant}: a predicate that expresses the properties of the shared variables that must be preserved in all the states visible by all the participating threads.
The general judgement in CSL, denoted as $\RI \vdash \hoaretriple{P}{
  C }{Q}$, expresses that with a resource invariant $\RI$, the
execution of $C$ satisfies the Hoare triple $\hoaretriple{P}{ C }{Q}$. 
The resource invariant can be obtained by executing operations of the
associated synchroniser. For example, a concurrent program
synchronised with a single-entrant lock can be verified as follows~\cite{GotsmanBC11}: any thread that successfully obtains the lock acquires $\RI$ and before releasing the lock it has to detach $\RI$ from its local state.
Verification of an atomic operation proceeds similar to verification
of a class using a lock: (1) the thread executing an atomic operation
acquires the global lock, (2) it adds the shared resources that are
captured by the resource invariant to its local state, (3) it performs
its action on the resources, (4) it establishes the resource
invariant, and finally, (5) it releases the lock by separating itself
from the resource invariant (and all this is done atomically). This is
described formally by the following rule of Vafeiadis~\cite{Vafeiadis11}:
\[
\frac
{
\emphp \vdash \hoaretriple{P \SLAnd \RI}{ C }{\RI \SLAnd Q}
}
{ 
\RI \vdash \hoaretriple{P}{\atomic C }{Q}
}
\quad
\rulename{Atomic}
\]
where $\RI$ is the resource invariant, $\emphp$ is the empty heap, ${\atomic C }$ indicates that
the command $C$ is executed atomically, $P$ is the precondition for
execution of the atomic operation, and $Q$ is the postcondition of the
atomic operation. 

To enable reasoning about multiple threads simultaneously reading the
same shared data, 
CSL has been extended with permissions~\cite{Boyland03} to
PBSL. This
extension is necessary to specify and verify \emph{shared-reading} synchronisations~\cite{BornatCOP05}.
In PBSL, any access to location of the heap is decorated with a fractional permission $\pi \in (0,1]$.
Any fraction $\pi \in (0,1)$ is interpreted as a \emph{read} permission and the full permission $\pi = 1$ denotes a \emph{write} permission (full ownership).
Permissions can be transferred between threads at synchronisation points (including thread creation and joining). 
A thread can only mutate a location if it has the write permission to
that location. 
Based on the following rule, permissions can be split and combined to change between read and write permissions:
$
\ppointsto {e}{\pi}{v} \SLAnd \ppointsto{e}{\pi'}{v} \Equiv \ppointsto {e}{\pi + \pi'}{v}
$ , where $\pi + \pi'$ is undefined if the result is greater than $1$.

Soundness of the logic ensures that the sum of
all permissions to a location is never more than 1. 
Thus, at most one thread at a time can be writing to a location, and whenever a thread has a
read permission, all other threads holding a permission on this location simultaneously must have a read permission. 
As a result, a verified concurrent program using PBSL is data-race free.
This makes PBSL to specify the behaviour of a shared-reading synchronisation mechanism.

In this paper we are using our \vercors specification
language to specify and verify the behaviour of synchronisers (in
\cref{sec:contracts}). The specification language of \vercors is an
extension of the Java Modeling Language (JML) with PBSL. The
standard SL notation of $\SLAnd$ for separating conjunction becomes
\incode{**} in our specifications, in order to avoid a syntactical
clash with the multiplication operator of Java (and JML).  Method and
class specifications can be preceded by a \incode{given} clause,
declaring ghost parameters to method and classes.  Ghost method
parameters are passed at method calls, ghost class parameters are
passed at type declaration and instance creation, resembling the
parametric types mechanism of Java. This mechanism is used to pass
resource invariants to classes.  Furthermore, the language has support
to declare abstract predicates~\cite{ParkinsonB08}, by providing the name, typing and
parameter declaration.

The full grammar for the \vercors specification language is as follows:
\[
\begin{array}{lrl}
R & ::= & B \mid \incode{Perm(field, pi)} \mid (\incode{\\forall* T i} ; B ; R )
\\ & \mid & R_1 \incode{**} R_2  \mid B_1 \incode{==>} B_2 \mid E. \incode{P} ( E_1 , \cdots, E_2 )
\\
E & ::= & \mbox{any {\em pure} expression}
\\
B & ::= & \mbox{any {\em pure} expression of type boolean}
\end{array}
\]
where $R$ denotes \emph{resource expressions} (typical elements $r_i$), $E$ represents \emph{functional expressions} (typical elements $e_i$), $B$ is the logical expressions of type boolean (typical elements $b_i$), 
\incode{T} is an arbitrary type, \incode{vi} is a variable name, \incode{P} is an abstract predicate of a special type \incode{resource}, \incode{field} is a field reference, and \incode{pi} denotes a fractional permission.

\section{Shared-reading Synchronisers} \label{sec:synchronisers}

In Java, volatile variables can be used as a communication mechanism between multiple threads. 
Writing to (or reading from) a volatile field has the same memory effect as if a monitor is released (or locked).
Therefore when writing to a volatile variable, its value becomes immediately visible to other threads. 
This guarantees that reading a volatile field, always gets the latest completed written value. 
This is an essential feature for synchronisers, because all threads must have a \emph{consistent view} on the state of the synchroniser. 

The \incode{atomic} package of JUC contains a set of atomic classes that define wrapper functions for private volatile fields with different types. 
Each atomic class defines three basic atomic operations. For example the \incode{AtomicInteger} class exports \incode{get()} for atomic read, \incode{set(int v)} for atomic write and \incode{compareAndSet(int x,int n)} for atomic conditional update.
The \incode{compareAndSet(int x,int n)} method first atomically checks the current value of the volatile field and updates it to \incode{n} if it is equal to the expected value \incode{x}, otherwise it leaves the state unchanged, and then it returns a boolean to indicate whether the update succeeded.
%
This \incode{AtomicInteger} class is the basis for almost all synchroniser implementations in \incode{java.util.concurrent},
such as \incode{ReentrantLock} and other classes implementing the interface \incode{Lock}, {\incode{Semaphore}},  \incode{CyclicBarrier} and \incode{CountDownLatch}.

Here we present (simplified) implementations of two different shared-reading synchronisation classes: \incode{Semaphore} and \incode{CountDownLatch}.
In our implementations, we stripped fairness conditions from the
original soure code, \ie we did not implement any algorithm to fairly pick the next candidate for the shared resource competition.
These examples illustrate how atomic variables are used in
shared-reading synchronisers, which will  help us to explain the formal specification in \cref{sec:specifications}. Finally, in~\cref{sec:verification} we will demonstrate how these synchronisers are verified.

\begin{figure}[t]
\begin{minipage}[t]{0.5\linewidth}
\begin{lstlisting}[numbers=left,basicstyle=\small\sffamily,caption={Semaphore: Implementation.},,label={lst:semaphoreimpl},escapechar=\#]]
public class Semaphore{
  private AtomicInteger sync;
  Semaphore(int n){ 
    sync = new AtomicInteger(n); }

  public void acquire(){
    boolean stop = false; int c = 0;
    while(!stop) {
      c = sync.get(); #\label{acquireget}#
      if( c > 0 ){
        int nextc = c-1;
        stop = sync.compareAndSet(c,nextc); #\label{acqcomp}#
      }
    }
  }	
  public void release(){
    boolean stop = false;
    while(!stop) {
      int c = sync.get(); #\label{releaseget}#
      int nextc = c+1;
      stop = sync.compareAndSet(c,nextc); #\label{relcomp}#
    }
  }
}
\end{lstlisting}
\end{minipage}
\hspace{0.5cm}
\begin{minipage}[t]{0.5\linewidth}
\begin{lstlisting}[numbers=left,basicstyle=\small\sffamily,caption={CountDownLatch: Implementation.},,label={lst:countdownlatchimpl},escapechar=\#]]
public class CountDownLatch{
  private AtomicInteger sync;
  CountDownLatch(int count){ 
    sync=new AtomicInteger(count); }
  
  void countDown(){
    boolean stop = false;
    int c = 0 , nextc = 0;
    while(!stop){
      c = sync.get(); #\label{countdownget}#
      if (c > 0){
        nextc = c-1;
        stop = sync.compareAndSet(c, nextc); 
      }
    }
  }
  void await(){
    int c = sync.get();
    while(c!=0) {  c = sync.get(); } #\label{cd-awaitget}#
  }
}
\end{lstlisting}
\end{minipage}
\end{figure}

In a \incode{Semaphore} (see~\cref{lst:semaphoreimpl}) all participating threads compete with each other to acquire or release protected portions of the shared resource.
In a concurrent program synchronised with a semaphore, any thread trying to acquire a portion, has to win the competition by atomically decrementing the number of available portions (see line~\ref{acqcomp} of~\cref{lst:semaphoreimpl}).
Similarly, as implemented in line~\ref{relcomp} a releasing thread (again in a competition) must atomically increment the number of available portions.

Next we consider a \incode{CountDownLatch}. 
Suppose we have an application with disjoint sets of active and passive threads, where active threads initially own a portion of the shared resource and passive threads wait for active threads to release their portions.
\incode{CountDownLatch}, as implemented
in~\cref{lst:countdownlatchimpl} blocks all the passive threads, until
all active threads have released their portion of the shared resource.
If the passive threads are unblocked, ownership of the shared resource
is transferred to the passive threads.

A~\incode{CountDownLatch} maintaints a counter that denotes the number of active threads working on the shared resource. 
Each active thread, once finished, calls~\incode{countDown()} on the
latch, which decreases the counter (see line~\ref{countdownget}), to
signal that it is done.
The passive threads wait for the active threads by calling the blocking~\incode{await()} method on the latch.
Inside this method, the passive threads are continuously reading the state of the latch until it reaches zero (line~\ref{cd-awaitget}).
In fact, the latch collectively accumulates the full shared resource from the active threads and the waiting passive threads can continue their task only when they see that there is no more active thread possessing a portion of the shared resource.

In summary, groups of threads involved in the synchronisation can be abstracted by their behavioral \emph{role}. 
If threads with an identical role share a resource (as in \incode{Semaphore}), then in order to obtain (or release) a portion of that resource, they have to participate in a \incode{compareAndSet}-based competition. 
But, if threads have different roles (as the passive and active thread groups in \incode{CountDownLatch}) they can exchange the shared resource by reading the atomic variable that controls the access. 
Note however, active threads in the \incode{CountDownLatch} still have to compete with each other to release their portions.
In both of these synchronisers, the state of the volatile counter defines the remaining portion of the shared resource.
Intuitively, associating the role of the threads, the state of the synchroniser and the portions of the shared resources distributed among the threads are the main elements in our reasoning about synchronisers which is explained in the next section. 

\section{Reasoning about Atomics} \label{sec:specifications} 

This section extends the formal specifications of the atomic
operations presented in~\cite{AmighiBH14} in such a way that they can
be used to verify both exclusive access and shared-reading
synchronisation constructs. 

We~\cite{AmighiBH14} identified various
synchronisation patterns using basic atomic operations. These
synchronisation patterns show that a thread:
(1) can both obtain or release resources by calling the \incode{compareAndSet} (or simply {\opcas})
operation, if it wins the competition, 
(2) may only obtain resources by calling the {\opget} operation,
provided it meets the conditions imposed by the protocol on the
thread's view of the atomic variable, and
(3) always releases resources by writing an atomic location using the
{\opset} operation. 

To explain the essence of our specification, first, we focus on competitive resource acquisition using the {\opcas} operation. We start with a simple example that illustrates the behavior of atomic variables to see how a fraction of the shared resource is exchanged when an atomic variable is used as a shared-reading synchronisation mechanism. 

Similar to the formalisation for exclusive access
synchronisers~\cite{AmighiBH14}, we partition the heap augmented with permissions into two disjoint parts, denoted $\ALocs$ for atomic locations and $\NLocs$ for non-atomic locations. 
For a given atomic variable $\avar s \in \ALocs$, we restrict the set of atomic operations to:
(1) $\opget(\avar s)$ for atomic read of $\avar s$, (2) $\opset(\avar s,\var n)$ for atomic update of $\avar s$ with the value $\var n$, and (3) $\opcas(\avar s,\var x,\var n)$ for conditional atomic update of $\avar s$ from the value $\var x$ to the value $\var n$. 
Resources are defined as locations from the non-atomic part of the heap.
Further, we extend the interval of the permissions to include $0$ and
we define $\ppointsto{e}{0}{-} \equiv \emphp$ ($\emphp$ denotes the empty heap).

As an example, using a semaphore $\avar s \in \ALocs$ to protect a location $\var r \in \NLocs$, the value of the atomic location $\avar s$ (defined as atomic state) indicates the number of available fractions for the semaphore.
The resource invariant for $\avar s$ associates the value of $\avar s$ with the maximum number of threads that concurrently can read $\var r$ and is defined as:
\[
\RI_{\avar s} =  
\quantify{\exists}{ v \in \set{0,\cdots,\var M}}
{
	\ppointsto{\avar s}{1}{v} 
	\SLAnd 
	\ppointsto{\var r}{\frac{\var v}{\var M}}{-}
}
\]

In an implementation of the semaphore, any thread that wishes to acquire a portion of the shared resource must atomically decrement the value of $\avar s$ by $1$.
This transfers $\frac{1}{\var M}$ of $\var r$ from $\avar s$ to the calling thread.
This fraction is stored back to $\avar s$ by releasing the semaphore, which increments the current value of $\avar s$ atomically by $1$.

In the implementations of \incode{acquire} and \incode{release}, the
executing thread with expected value $\var x$ executes the atomic body
of the $\opcas$ operation. As justified by the atomic rule, to verify
the body it obtains $\RI_{\avar s}$. This gives full access of $\avar
s$, as well as $\frac{\var x}{\var M}$ of $\var r$
(provided  the current state equals $\var x$). 
The thread then updates $\avar s$ to $\var n=x-1$ for \incode{acquire} or $\var n=x+1$ for \incode{release}, and re-establishes $\RI_{\avar s}$ with $\ppointsto{\var r}{\frac{\var n}{\var M}}{-}$ before leaving the body.
To do so, the thread either acquires itself a $(\frac{\var x}{\var
  M}-\frac{\var n}{\var M})$ fraction of $\var r$ or it releases a $(\frac{\var n}{\var M}-\frac{\var x}{\var M})$ fraction of $\var r$. 
This example gives us  the necessary intuition to derive a
specification for the {\opcas} operation to cover both \emph{shared and exclusive} synchronisers.

If we denote the shared resources to be protected by the atomic location $\avar s$ using $\resmap(\avar s)$, then we can define the resource invariant as:

\[
\begin{array}{lr}
\RI_{\avar s} 

= 
\quantify{\exists}{ v \in \set{0,\cdots,\var M}}
{
	\ppointsto{\avar s}{1}{v} 
	\SLAnd 
	S(\avar s,\pi)
}
\end{array}
\]
where
$
S(\avar s,\pi) =
\fastar{r \in \resmap(\avar s)}{ \ppointsto{r}{\pi}{-} }
$ .

Using $\RI_{\avar s}$, the atomic location $\avar s$ is interpreted as
the owner of the resources for which the threads compete through the
{\opcas} operation in order to obtain or release their permissions. 
Based on this general definition of resource invariant, we can specify the behavior of {\opcas}. 
For a synchroniser $\avar s$, if $\fracfun$ maps the state of the synchroniser to the fractions with a maximum number of threads $M$, then we can axiomatise {\opcas} as follows:
\[
\frac
{
\pi=\fracfun(\avar s, \var x, \var M) \quad \pi'=\fracfun(\avar s,\var n, \var M) \quad b=\opcas_{\tid}(\avar s,\var x,\var n)
}
{ 
\RI_{\avar s} \vdash 
\begin{array}{l}
\outline{S(\avar s,\pi' \fracminus \pi)} 
\\
{\opcas_{\tid}(\avar s,\var x,\var n)}
\\
\outline{(b \implies S(\avar s,\pi \fracminus \pi')) \SLAnd (\neg b \implies S(\avar s,\pi' \fracminus \pi))}
\end{array}
}
\
\rulename{CAtm}
\]
where $\fracminus$ denotes the cut-off subtraction over the fractions
in $[0,1]$, defined as follows: 
\[
\pi \fracminus \pi' = 
\left\{
\begin{array}{ll}
\pi - \pi' & \text{iff} \ \pi \geq \pi', 
\\
0 & \text{otherwise}
\end{array}
\right.
\]

Surprisingly, the behaviors of both atomic read and write are more
subtle than for the {\opcas} operation. 
This is because their
behavior can differ from one case to another. In some cases, the
atomic read operation only updates the knowledge of the executing
thread without transferring any resources: see lines~\ref{acquireget}
and~\ref{releaseget} in~\lstlistingname~\ref{lst:semaphoreimpl},
line~\ref{countdownget}
in~\lstlistingname~\ref{lst:countdownlatchimpl}. Also, the waiting
threads in \incode{CountDownLatch} (see line~\ref{cd-awaitget} from
\cref{lst:countdownlatchimpl}) obtain their fractions only when they
realize that the latch has reached zero.  In other cases,
unconditional updates in the atomic writes require a
\emph{rely-guarantee}~\cite{Jones83-ifip} style of reasoning as the
writing thread must adhere to a \emph{protocol} which guarantees the
safety of the write to the environment~\cite{AmighiBH14}.  This is
thoroughly discussed and formalised by Amighi \etal \cite{AmighiBH14}. Here we extend the formal definition of the
resource invariant from~\cite{AmighiBH14} to associate the state of
the atomic variable with the fractions of the resources. First, we
explain some notations which are used in the definitions.

A thread view is an \emph{atomic ghost field} defined for each thread that stores the last visited value of the atomic state.
Each view is indexed by the owning thread identifier and the ownership of a view is split between the owning thread and the resource invariant, thus, it can only be updated inside an atomic body.
$\vector{\viewofs t}$ denotes the vector of views, indexed by their thread identifiers.
A vector of values pointed to by the views, indexed by the corresponding thread identifiers, is written  $\vector{\var v_t}$, while $\vector{\var v_t}_{\set{v_{\tid}=x}}$ denotes a vector such that the value of the item indexed with $\tid$ in the vector of values $\vector{\var v_t}$ is equal to $x$. Finally, having defined the views of the threads the synchronisation construct is  generalised from a single atomic location $\avar s$ to a tuple of the atomic location plus all the thread views of this location.

We decomposed the resource invariant into two components~\cite{AmighiBH14}. 
The first component is the global resource invariant that associates the resources to the \emph{global} atomic state: 
\[
 \RI_{\avar s} = 
 	\exists \var v, \vector{\var w_t} \in \States \cdot
 	{  \fullpto{\avar s}{\var v} 
 	    \SLAnd
 	    (\fastar{t \in \Threads }{ \halfpto{\viewofs t}{\var w_{t}} })
 	    \SLAnd 
 		S(\avar s, \pi)  
 	} 
 	\SLAnd \protocol^{\Threads}_{\avar s}
\]
where:
\begin{itemize}  
\item having $\feasible$ for determining the feasibility of the values taken by the atomic location and all the thread views $\protocol_{\avar s}^{\Threads}$ is defined as follows:
\[
\protocol_{\avar s}^{\Threads} = 
	\underset{{\var v,\vector{w_t} \in \States} \cdot {\feasible(\var v,\vector{w_t})}}{\bigvee}(\astate {\avar s}=v \AND \vector{\astate{\viewofs t}} =\vector{\var w_t})
\]

\item the fraction of the resources is associated with the atomic state via $\pi = \fracfun(\avar s,\var v, \var M)$, and
\item finally, $
S(\avar s, \pi)  =\fastar{r \in \resmap(\avar s)}{ \ppointsto{r}{\pi}{-} }$

\end{itemize}

The second component associates the fractions of the resources to the thread views which can be exchanged through a collaborative synchronisation: 
\[
T(\avar s_t, \pi)  =
\fastar{r \in \resmap(\avar s_t)}{ \ppointsto{r}{\pi}{-} }
\ \text{ where } \
\pi = \fracfun(\avar s_t,\var w_t, \var M)
\]

By giving a definition for $T(\avar s_t, \pi)$ one can express when a thread with a particular knowledge may obtain the resource.
The {\opset} absorbs the resources either through $\RI_{\avar s}$ to the atomic location or through $T(\avar s_t, \pi)$ to the reader thread.
This is formally specified in the contracts for the basic atomic operations which are presented in ~\cref{fig:spec-final}.

Comparing the new contribution with \cite{AmighiBH14}, our formalised
extension for shared-reading synchronisers can be summarised by the
following steps: (1) an extension of the permission interval with $0$,
(2) associating the fractions of the shared resource to the global atomic state, (3) and updating the contract of {\opcas} using the cut-off subtraction operation.  

The next section presents how the specification from~\cref{fig:spec-final} translates into a contract for the {\incode{AtomicInteger}} class, using our \vercors \cite{vct} specification language. 

\begin{figure*}[t]
{\compsize
\[
\begin{array}{l}
 \frac
 	{ 
\begin{array}{l}
	 	\pi=\fracfun(\avar s, \var n, \var M) \quad \pi'=\fracfun(\viewofs{\tid},\cnst d, \var M) \quad \quad
 		\quantify{\forall}{ \var v,\vector{\var v_t} \in \States}{\var v_{\tid}=\cnst d \AND \feasible(\var v,\vector{\var v_t}_{\set{\var v_{\tid}=\cnst d}})} \implies \feasible(\var n,\vector{\var v_t}_{\set{\var v_{\tid}=\var n}})
 \end{array}
	} 
 	{
\begin{array}{ll}
 \RI_{\avar s} \vdash
&
\htriple
	{ 
	\halfpto{\viewofs{\tid}}{\cnst d} \SLAnd \sresz{\avar s}{\pi} \SLAnd \tresz{\viewofs{\tid}}{\pi'}
	}
	{
		\opset_{\tid}(\avar s,\var n)
	}
	{ 
	\halfpto{\viewofs{\tid}}{\var n} 
	}
\end{array}
	}
	\quad
	\rulename{WAtm}
\\
[20pt]
\\
\frac
	{
	\begin{array}{c}
	 	\pi=\fracfun(\viewofs{\tid}, \var M) \quad \pi'=\fracfun(\viewofs{\tid},\var M)
	\end{array}	
	}
	{
\begin{array}{ll}
 \RI_{\avar s} \vdash
&

\htriple
	{ 
	\halfpto{\viewofs{\tid}}{\cnst d} \SLAnd \tresz{\viewofs{\tid}}{\pi}
	}
	{ 
		\opget_{\tid}(\avar s) 
	}
	{ 
	\halfpto{\viewofs{\tid}}{\var \ret} \SLAnd \tresz{\viewofs{\tid}}{ \pi'}
	}
\end{array}
	}
	\quad
	\rulename{RAtm}
\\
[20pt]
\\

\frac 
	{
\begin{array}{l}
 		\pi=\fracfun(\avar s, \var x, \var M) \ \ \pi'=\fracfun(\avar s,\var n, \var M) \ \
 \\		
 		\quantify{\forall}{ \var v,\vector{\var v_t} \in \States}{\var v_{\tid}=\var x \AND \feasible(\var v,\vector{\var v_t}_{\set{\var v_{\tid}=\var  x}})} \implies \feasible(\var n,\vector{\var v_t}_{\set{\var v_{\tid}=\var n}})
 		\ \
 	b=\opcas_{\tid}(\avar s,\var x,\var n)
\end{array}
	}
	{
\begin{array}{lll}
\RI_{\avar s} & \vdash &
\begin{array}{l}
\outline{	\halfpto{\viewofs{\tid}}{\var x} \SLAnd \sresz{\avar s}{\pi' \fracminus \pi} }
\\
{\opcas_{\tid}(\avar s,\var x,\var n)}
\\
\outline{ 
	(b \implies \halfpto{\viewofs{\tid}}{\var n} \SLAnd \sresz{\avar s}{\pi \fracminus \pi'}  ) 
		 \OR  
	(\neg b \implies \halfpto{\viewofs{\tid}}{\var x} \SLAnd \sresz{\avar s}{\pi' \fracminus \pi}  
}
\end{array}
\end{array}
	}
	\quad
	\rulename{CAtm}
\end{array}
\]
 }
\caption{Thread-modular specifications of atomic operations} \label{fig:spec-final}
\end{figure*}

\section{Contract of Atomic Integer} \label{sec:contracts}

The new contract of \incode{AtomicInteger} is presented in \cref{lst:SharedAtomicIntegerSpecification}. 
First we summarise the elements of the contract that are defined from
our earlier work \cite{AmighiBH14}. Then, we explain our extensions regarding shared-reading synchronisers.

\begin{lstlisting}%
[float=t, numbers=left,basicstyle=\small\sffamily, caption={Contracts for \incode{AtomicInteger}: Exclusive and Shared-reading},label={lst:SharedAtomicIntegerSpecification}]
/*@given Set<role> rs; #\label{ai:rolesdef}# 
given group (frac->resource) inv; #\label{ai:invdef}#   
given (role,int->frac) share; #\label{ai:sharedef}# 
given (role,int,int-> boolean) trans; #\label{ai:transdef}#  @*/
class AtomicInteger {    
   private volatile int value;
/*@group resource handle(role r,int d,frac p); #\label{ai:handle}#  @*/
/*@requires inv(share(S,v));    ensures (\forall* r in rs: handle(r,v,1));  @*/
  AtomicInteger(int v);
  
/*@given role r, int d, frac p;
requires handle(r,d,p) ** inv(share(r,d));
ensures handle(r,\result,p) ** inv(share(r,\result));  @*/
  public int get();
	
/*@given role r, int d, frac p;
requires handle(r,d,p) ** trans(r,d,v);
requires inv(share(S,v)) ** inv(share(r,d)); 
ensures handle(r,v,p);@*/  
  public void set(int v);
	
/*@given role r, int m, frac p;
requires handle(r,x,p) ** trans(r,x,n) 
requires inv(share(S,n)-share(S,x));
ensures  \result==> (handle(r,n,p) ** inv(share(S,x) - share(S,n)); 
ensures !\result==> (handle(r,x,p) ** inv(share(S,n) - share(S,x));  @*/  
  boolean compareAndSet(int x, int n);
}
\end{lstlisting}

\cref{lst:SharedAtomicIntegerSpecification} shows how the
\incode{AtomicInteger} class is parametrised by \incode{rs},
\incode{inv}, \incode{share}, \incode{trans}, where
\incode{rs} is a set of roles abstracting participating threads, 
\incode{inv} represents an abstract predicate as a resource invariant,
specifying the shared resources to be protected by
\incode{AtomicInteger}, and 
\incode{share} defines a function to associate the states of the atomic integer with a fraction of the shared resource; and 
\incode{trans} is a boolean predicate, encoding all the valid transitions that a particular instance of \incode{AtomicInteger} can take.

An instance of \incode{AtomicInteger} as the coordinator of threads is specified using a globally known role \incode{S}.
Any thread calling a method of the \incode{AtomicInteger} can acquire
or release a fraction of the shared resource, depending on its role
and the current state of the atomic integer. This is specified in
\incode{AtomicInteger}s contract. 
In order for a thread to be eligible to call a method, it has to
possess a token indicating its role and the value of the atomic state
upon its  last visit.
This is captured by an abstract predicate \incode{handle} (line~\ref{ai:handle} of~\lstlistingname~\ref{lst:SharedAtomicIntegerSpecification}).
Essentially, this abstract predicate witnesses the role of the calling
thread, its last seen  value of \incode{AtomicInteger}, and the fraction of the token.
 
The constructor of the \incode{AtomicInteger} absorbs the resource associated with the initial value.
Often, as  seen e.g. in \incode{Semaphore}, if the resource is obtained by a \incode{compareAndSet}-based competition, then the synchroniser owns the resource at the beginning.
But, if the threads start their life with some resources in their
hands (such as the active threads in \incode{CountDownLatch}), then the synchroniser does not own any resource in its initial step.

The \incode{get} method exchanges the resources based on the view of the calling thread. 
In a competition-based synchronisation threads do not obtain any
resource by calling the \incode{get} method;  they only update their
knowledge about the current state.
Apart from the thread's handle, any thread calling \incode{set(int v)}
has to provide (1) its permission to write the value \incode{v} to the
atomic integer, (2) the resources associated to its current view, and
(3) the resources associated to the value \incode{v}, \ie the next state of the atomic integer. 
Upon return of the \incode{set} method, the calling thread only obtains a handle updated with the thread's new view. 
Also the thread trying to atomically update the value of an atomic integer
by calling \incode{compareAndSet(int x,int n)}  has to have the
permission for the transition from \incode{x} to \incode{n} and the
right handle to call this operation.

Following our formal specification (see~\cref{fig:spec-final}), our
extension of the contract of \incode{AtomicInteger} captures that the \incode{compareAndSet(int x, int v)} method absorbs the \emph{difference} between the resources that the synchroniser will hold in case of a successful update, \ie the resources associated with \incode{n}, and the resources that the synchroniser object currently holds, \ie the resources associated with \incode{x}. 
If the operation succeeds, the operation ensures the difference between the resources that the synchroniser owned before the call, \ie the resources associated on \incode{x}, and the resources that holds after the successful update, \ie the resources associated with \incode{n}.
If the operation fails, no resources are exchanged. Instead all resources specified in the pre-condition are returned.
In the specification of \incode{AtomicInteger}, the difference between
the resources turns into the subtraction operation between two fraction types.
The subtraction between two permissions is defined as zero if the result of the operation becomes negative.
Besides, as explained above, \incode{inv(0)} is equivalent to \incode{true}.
Therefore, as expressed in the contract of \incode{compareAndSet}, the difference between the resources associated with the two states \incode{x} and \incode{n} determines if the calling thread releases or obtains fractions of the shared resource.

Essentially, our extension for the contract of \incode{AtomicInteger} class is realised by: (1) defining the \incode{share} function to map the fractions to the atomic state, and (2) updating the specification of \incode{compareAndSet} with the cut-off subtraction between fractions. 
In the next section we demonstrate how one can use this specification
of \incode{AtomicInteger} to verify an implementation of a
shared-reading synchroniser.

\section{Verification} \label{sec:verification}
In this section we demonstrate how to verify the specification of a shared-reading synchroniser \wrt its implementation using an instance of \incode{AtomicInteger}.
For space reasons, we only explain the verification of
\incode{Semaphore} using the \vercors tool set; the verification of \incode{CountDownLatch} is
very similar to \incode{Semaphore}. Moreover, to show that our new
specification still supports verification of exclusive access
synchronisers, we have also verified an implementation of a
\incode{SpinLock}. All examples are
available online~\cite{vct-lock, vct-semaphore, vct-countdownlatch}.
The examples are automatically verified using  the \vercors tool set~\cite{vct}. \vercors is the tool that encodes our specified programs to intermediate languages like Viper~\cite{0001SS16} and Chalice~\cite{LeinoMS09} to be verified by permission-based SL back-ends like Silicon~\cite{0001SS16}.


\subsection{Semaphore: verification}
Class \incode{Semaphore} implements a synchroniser where a group of threads simultaneously can have read access to a shared resource.
The full code for this class is specified in~\cref{lst:semaphoreverification-cons},~\cref{lst:semaphoreverification-acquire} and~\cref{lst:semaphoreverification-release}.

The \incode{Semaphore} class is parametrized with the resource invariant defined by its client program.
The instantiated semaphore uses two predicates as tokens to detect if a thread holds a fraction of the shared resource: \incode{initialized} and \incode{held}. 

An instance of a semaphore protects a shared resource with a specified maximum number of permits which is defined as a ghost variable within the class (line~\ref{sem:ghostdef} of \cref{lst:semaphoreverification-cons}).
To instantiate an object of \incode{AtomicInteger}, the
\incode{Semaphore} class has to define the required
protocol. Resources are acquired using a compare-and-set based competition.
All participating threads have identical roles in the specification.
The shared resource to be protected by \incode{AtomicInteger} is the
same as the surrounding program passes on to the semaphore (\cref{lst:semaphoreverification-cons}, line~\ref{sem:invdef}).
The definition of the \incode{share} function defines the fraction of the shared resource that must be held by \incode{AtomicInteger} in each state (\cref{lst:semaphoreverification-cons}, line~\ref{sem:sharedef}).
The definition given for the valid transitions expresses that in each update the difference between two states must be one unit (\cref{lst:semaphoreverification-cons}, line~\ref{sem:transdef}).

\subsubsection{Constructor }

\begin{lstlisting}[float=t,numbers=left, basicstyle=\small\sffamily , caption={Verification of \incode{Semaphore}: constructor.},label={lst:semaphoreverification-cons},escapechar=\#]
/*@ given group (frac -> resource) rinv; #\label{sem:rinvdef}# @*/
public class Semaphore{
  /*@ ghost final int num;  #\label{sem:ghostdef}#   ghost Set<role> roles = {T}; #\label{sem:roledef}#
  group resource initialized(int d,frac p) = sync.handle(T,d,p);  #\label{sem:initializeddef}#
  resource held(int d,frac p) = initialized(d,p);
  group resource inv(frac p) = rinv(p);  #\label{sem:invdef}#
  frac share(role r, int c){  #\label{sem:sharedef}# return (r==S && c>=0 && c<num)?(c/num):0; } 
  boolean trans(role r, int c, int n){  #\label{sem:transdef}# 
      return (r==T && c>0 && n==c-1) || (r==T && c<max && n==c+1); } @*/
  private AtomicInteger/*@ <roles,inv,share,trans> @*/ sync;

  /*@ requires rinv(1) ** n>0;  #\label{sem:consreq}#  ensures initialized(n,1) ** num == n;  #\label{sem:consens}# @*/  
  Semaphore(int n){
  /*@ set num = n;  fold sync.inv(share(n)); #\label{sem:confold}#  @*/
     sync=new AtomicInteger/*@<roles,inv,share,trans>@*/(n);
  /*@ fold initialized(n,1); @*/
  }
}
\end{lstlisting}

The client of the semaphore instantiates the object with a number of available units to acquire.
Thus, it has to provide the resources associated with the initial value of the semaphore.
After storing the maximum number of permits in the ghost field \incode{num}, the body of the constructor can feed the \incode{AtomicInteger} class with the resources associated with its initial value (see line~\ref{sem:confold} of \cref{lst:semaphoreverification-cons}).
In return, the constructor of \incode{AtomicInteger} returns its handle, which can be used to establish the postcondition of the constructor of the semaphore as defined in line~\ref{sem:initializeddef}.
Finally, the semaphore ensures a full initialized token to the client
program, which can be distributed in portions among the participating threads.   

\subsubsection{Methods}

The annotated versions of the methods \incode{acquire()} and \incode{release()} are presented in~\cref{lst:semaphoreverification-acquire} and~\cref{lst:semaphoreverification-release}, respectively. Having a fraction of the initialized token given by the client program, each thread is authorized to start its competition to acquire a permit of the shared resource protected by the semaphore.
First, the acquiring thread has to read the current state of the atomic integer to see how many permits are still available. To achieve this, the body of the acquire has to unfold the provided initialized token to capture the required handle of the \incode{get} method from \incode{AtomicInteger} (line~\ref{sem:acqunfold} of ~\cref{lst:semaphoreverification-acquire}).
According to the provided protocol for the \incode{AtomicInteger}, the thread does not have any resource associated with its view.
Therefore, having the right handle suffices to read the current state of the \incode{sync} object (see line~\ref{sem:getcall} of~\cref{lst:semaphoreverification-acquire}).
To acquire a unit of the available permits the thread must decrement the current state by one.
So it folds all the required abstract predicates as the specification of \incode{compareAndSet} demands.
Based on the given definition for the protocol, the acquiring thread does not need to provide any resources at this step.
In case of a successful update, the \incode{compareAndSet(c,nextc)} returns one unit of the shared resource, \ie \incode{inv(1/num)} (see~\ref{sem:acqsuccas} of~\cref{lst:semaphoreverification-acquire}).
The successful thread can then leave the body of \incode{acquire} after folding the \incode{held} predicate using the available handle from \incode{AtomicInteger}.
In the post-condition of the \incode{acquire} method, \incode{?w} denotes the existence of a view for the calling thread after the call.
Finally, if the thread fails to decrement the state, it continues reading the current state and trying to atomically decrement the state.

\begin{lstlisting}[float=t,numbers=left, basicstyle=\small\sffamily, caption={Verification of \incode{Semaphore::acquire()}.},label={lst:semaphoreverification-acquire},escapechar=\#]
/*@ given int d, frac p;
requires initialized(d, p) ** d<=num ** d>0;  #\label{sem:acqreq}#
ensures held(?w,p) ** rinv(1/num) ** w<num ** w>=0;#\label{sem:acqsens}#@*/  
  public void acquire(){
/*@ unfold initialized(d,p);  #\label{sem:acqunfold}# @*/
    boolean stop = false; int c = 0;
    while(!stop) { /*@ fold sync.inv(sync.share(T,d)); @*/
      c = sem.get(); #\label{sem:getcall}#
      if( c > 0 ){    int nextc = c-1;
  /*@ fold sync.trans(T,c,nextc);
  fold sync.inv(sync.share(S,nextc)-sync.share(S,c)); @*/
        stop = sem.compareAndSet(c,nextc); #\label{sem:acqsuccas}#
      }
    }  /*@ fold held(nextc,p); @*/
  }
\end{lstlisting}

Releasing a fraction of the shared resource is symmetric to the \incode{acquire} method. 
It should be easy to follow the reasoning steps presented in~\cref{lst:semaphoreverification-release}.
We only note here that the thread calling the \incode{release} method provides the fraction of the shared resource it owns.
Then, in an attempt to increment the current state of the atomic
integer, if it succeeds it gives up the permit by folding the \incode{inv} abstract predicate required by \incode{compareAndSet(c,nextc)} at line~\ref{sem:relcas} of \lstlistingname~\ref{lst:semaphoreverification-release}.

\begin{lstlisting}[float=t, numbers=left,basicstyle=\small\sffamily,caption={Verification of \incode{Semaphore::release()}.},label={lst:semaphoreverification-release},escapechar=\#]
/*@ given inr d,frac p;
requires held(d,p) ** rinv(1/num) ** d<num ** d>=0;  #\label{sem:relreq}#
ensures initialized(?w,p) ** w<=num ** w>0 ;  #\label{sem:relens}# @*/  
  public void release(){
/*@ unfold held(d,p); unfold initialized(d,p); @*/
    boolean stop = false;
    while(!stop) {
      int c = sync.get();       int nextc = c+1;
/*@ fold sync.trans(T,c,nextc); @*/
/*@ fold sync.inv(sync.share(S,nextc)-sync.share(S,c)); @*/
      stop = sync.compareAndSet(c,nextc);  #\label{sem:relcas}#
    } /*@ fold initialized(nextc,p); @*/
  }
\end{lstlisting}

\section{Conclusion and Related Work}\label{sec:conclusion}

Many different extensions of CSL are proposed in the literature. 
After RGSep~\cite{VafeiadisP07} and Deny-Guarantee Reasoning~\cite{DoddsFPV09}, 
CAP~\cite{Dinsdale-YoungDGPV10} was introduced to reason about atomic operations. In CAP, resources are encoded together with the environment interference in an atomic rule to reason about synchronisation with finer granularity.
The dream of having an universal logic for concurrent programs resulted in developing various extensions of CAP, namely HOCAP~\cite{SvendsenBP13}, iCAP~\cite{SvendsenB14} and, finally, Iris~\cite{JungSSSTBD15}. Iris is a PBSL based logic to reason about fine-grained concurrent data structures. It supports resource algebras, invariants and higher-order predicates. 
The user has to instantiate the logic with the elements of the target programming language. 
Currently, Iris-based verification is performed in Coq.
Finally, Caper~\cite{Dinsdale-YoungP17} is a verification tool where the core logic is based on CAP with additional features taken mainly from iCAP and Iris. 

All the above mentioned works focus on the development of a generic, universal and powerful program logics. 
Instead, we treat reasoning about atomic operations at the specification level using an already existing logic, \ie PBSL.
We reuse an existing specification language (JML) to support more intuitive assertion language.
This allows one to employ currently existing PBSL verifiers like Silicon~\cite{0001SS16} and Verifast~\cite{JacobsSPVPP11}.
We modified our approach in such a way that
the new specification of \incode{AtomicInteger} can be used to verify both exclusive and shared-reading synchronisers. 
This is done by defining a function that associates the atomic state to the fractions of the shared resources. 
The definitions of protocols and resource invariant are updated accordingly.
Then, by proposing the cut-off subtraction operation in permissions we updated the specifications of atomic operations.
We presented a set of mechanically verified examples to demonstrate how our new
specifications can be used to verify implementations of synchronisers.
In a separate work, we have presented how the specification of these synchronisation classes are used to verify a complete Java program~\cite{AmighiBDHMZ14}.




\section{Acknowledgments}
The work presented in this paper is supported by ERC grant 258405 for the VerCors project.

\bibliographystyle{eptcs}
\bibliography{vercors}  
\end{document}

%% file: defs-sac18.tex
\definecolor{dkgreen}{rgb}{0,0.6,0}
\definecolor{gray}{rgb}{0.5,0.5,0.5}
\definecolor{mauve}{rgb}{0.58,0,0.82}
\definecolor{reqcol}{rgb}{0.3,0.3,1}
\definecolor{enscol}{rgb}{0.3,0.3,1}
\definecolor{speccol}{RGB}{0,0,255}
\definecolor{faultyspeccol}{RGB}{205,0,0}
\definecolor{proofhintcol}{RGB}{20,140,70}

\newcommand{\tjkw}[1]{\mbox{\texttt{#1}}} 
\newcommand{\jkw}[1]{\tjkw{#1}} 

\newcommand{\setname}[1]{\mathsf{#1}}
\newcommand{\funname}[1]{\textsf{#1}}
\newcommand{\rulename}[1]{[\textsc{#1}]}
\newcommand{\ie}{\textit{i.e.}\xspace}

\newcommand{\etal}{\textit{et~al.}}

\newcommand{\wrt}{w.r.t.\xspace}

\def\lstlistingname{Listing~} 
\newcommand{\Threads}{\setname{Thr}}
\newcommand{\States}{\setname{Val}}

\newcommand{\ALocs}{\setname{ALoc}}
\newcommand{\NLocs}{\setname{NLoc}}

\newcommand{\disjoint}{\funname{disjoint}}

\newcommand{\feasible}{\funname{fsbl}}

\newcommand{\set}[1]{\{#1\}}

\newcommand{\resformula}[1]{\mathsf{#1}}

\newcommand{\OR}{\vee}                          
\newcommand{\AND}{\wedge}                          
\newcommand{\Equiv}{\Leftrightarrow}

\newcommand{\LLImplies}{\ensuremath{\tjkw{\raisebox{.08ex}{{-}}\kern-.09em*}}}
\newcommand{\LLEquiv}{\ensuremath{\tjkw{*\kern-.1em\raisebox{.02ex}{--}\kern-.1em*}}}
\newcommand{\SLAnd}{\ensuremath{\mathop{\tjkw{*}}}}

\newcommand{\pointsto}[2]{{#1}{\mapsto}{#2}}
\newcommand{\ppointsto}[3]{{#1}\stackrel{#2}{\mapsto}{#3}}
\newcommand{\bigslstar}{\circledast}
\newcommand{\emphp}{\resformula{emp}}

\renewcommand{\vector}[1]{\overset{\rightarrow}{#1}}

\newcommand{\cnst}[1]{\textsf{#1}} 
\newcommand{\var}[1]{#1} 

\newcommand{\avar}[1]{\var #1} 
\newcommand{\astate}[1]{[#1]} 

\newcommand{\RI}{I} 
\newcommand{\resmap}{\funname{res}}

\newcommand{\tresz}[2]{ \funname{T}(#1,#2)} 
\newcommand{\sresz}[2]{ \funname{S}(#1,#2) } 

\newcommand{\protocol}{\funname{P}}

\newcommand{\viewofs}[1]{\ensuremath{s_{#1}}}
\newcommand{\tid}{\tau}

\newcommand{\hoaretriple}[3]{\outline{#1} \ #2 \ \outline{#3}}

\newcommand{\result}{\textsf{result}}

\newcommand{\quantify}[3]{#1 \ #2.\ #3}

\newcommand{\outline}[1]{{\color{blue} \{#1\}}}

\newcommand{\ret}{ret}

\newcounter{rule}

\makeatletter
  \newcommand{\addToLabel}[1]{%
    \protected@edef\@currentlabel{\@currentlabel#1}%
  }
\makeatother

\newcommand{\opget}{\jkw{get}}
\newcommand{\opset}{\jkw{set}}
\newcommand{\opcas}{\jkw{cas}}

\newcounter{assertion}

\newcommand{\fullpto}[2]{\ppointsto{#1}{1}{#2}}

\newcommand{\halfp}{\frac{1}{2}}
\newcommand{\halfpto}[2]{\ppointsto{#1}{\halfp}{#2}}

\newcommand{\fastar}[1]{\underset{#1}{\bigslstar}}

\newcommand{\compsize}{\small}

\newcommand{\atomic}[1]{\ensuremath{\mathsf{atomic}\{\ #1\ \}}\xspace}

\newcommand{\pre}[1]{\ensuremath{{#1}_\mathit{pre}}}

\newcommand{\product}[2]{\ensuremath{\Pi_{#1}{#2}}\xspace}

\newcommand{\htriple}[3]{\outline{#1} \ #2 \ \outline{#3}}

\newcommand{\incode}[1]{\text{\lstinline+ #1 +}}

\definecolor{hintcol}{RGB}{255,255,100}

\newcommand{\vercors}{{\textsc{VerCors }}}

\crefname{section}{Section}{}
\crefname{chapter}{Chapter}{}
\crefname{lstlisting}{Listing}{}
\crefname{figure}{Figure}{}
\crefname{line}{Line}{}
\crefname{example}{Example}{}
\crefname{definition}{Definition}{}